\documentstyle [12pt]{article}

\begin{document}

\begin{titlepage}

\title{No-hidden-variables proof for two spin-$\frac{1}{2}$ particles preselected 
and postselected in unentangled states\footnote{Phys. Rev. A 
{\bf 55}, 4109 (1997).}}

\author{Ad\'{a}n Cabello\thanks{Electronic address: fite1z1@sis.ucm.es}\\
{\em Departamento de F\'{\i}sica Te\'{o}rica,}\\{\em Universidad Complutense, 
28040 Madrid, Spain, and}\\
{\em Departamento de F\'{\i}sica Aplicada,}\\{\em Universidad de Sevilla, 
41012 Sevilla, Spain.}}
\date{August 11, 1996}
\maketitle

\begin{abstract}
It is a well-known fact that all the statistical predictions 
of quantum mechanics on the state 
of any physical system represented by a two-dimensional Hilbert 
space can always be duplicated by a noncontextual hidden-variables model. 
In this paper, I show 
that, in some cases, when we consider an additional independent (unentangled) 
two-dimensional system, the quantum 
description of the resulting composite system cannot be reproduced using 
noncontextual hidden variables. In particular, a no-hidden-variables proof is 
presented for two 
individual spin-$\frac{1}{2}$ particles preselected in an uncorrelated state 
$\left| A \right\rangle \otimes \left| B \right\rangle $
and postselected in another uncorrelated state
$\left| a \right\rangle \otimes \left| B \right\rangle $,  
$\left| B \right\rangle $ being
the same state for the second particle in both 
preselection and postselection.

\end{abstract}

\end{titlepage}

Noncontextual hidden-variables (NCHV) models that are capable of reproducing all 
statistical predictions of quantum mechanics (QM) for physical systems described 
by {\em two-dimensional} 
Hilbert spaces do exist \cite{Bell66,BohmBub66,KS67,Clauser71,Selleri90}. 
Examples of such systems, 
also known as two-state systems or ``qubits'' \cite{Schumacher95}, include 
a single spin-$\frac{1}{2}$ particle without translational motion, 
the polarization of a 
photon, the relative 
phase and intensity of a single photon in two arms of an interferometer,
or an arbitrary 
superposition of two atomic states. On the contrary, for 
physical systems described by Hilbert 
spaces of dimension greater than 2, a fundamental theorem proved by Gleason 
\cite{Gleason57}, 
Bell \cite{Bell66}, and Kochen and Specker \cite{KS67} excludes 
NCHV alternatives to QM.  
On the other hand, Bell's theorem \cite{Bell64} prohibits local hidden variables 
(a particular type of NCHV) for {\em composite} systems of two (or more) parts 
(usually two-state systems) initially 
prepared in an {\em entangled} state. However, for systems composed 
of several uncorrelated 
(unentangled) two-dimensional subsystems, one might think that
NCHV descriptions are  possible. 
In this paper, I show that even in such a case, some quantum inferences 
cannot be duplicated using a
NCHV theory. For this purpose I present a simple no-go proof 
for an individual system of two spin-$\frac{1}{2}$ 
particles preselected and postselected \cite{ABL64} in uncorrelated states. 
The argument contains both measured (i.\ e., actual) and non-measured (i.\ e., 
hypothetical) values 
of the composite system. The former are the results of separate measurements
 on each particle 
in the preselection or postselection processes. 
The latter are hypothetical values of the whole system that are assumed to 
be determined 
(in a NCHV theory) in the time interval between the preselection and postselection, 
invoking one of the following criteria: (a) they can be predicted 
with certainty after the preselection; (b) they can be {\em retrodicted} 
\cite{ABL64,Vaidman93}
with certainty before the postselection; or (c)
they must verify the 
\mbox{{\em sum rule} \cite{Redhead87}} for the results of any measurement of 
an orthogonal resolution of 
the identity. The joint use of the preselection and postselection 
and of these three criteria to infer the values of some properties of the system
is legitimate in the context of a NCHV theory in which quantum observables
are assumed to have preexisting values revealed by the act of measurement (although 
of course not in QM itself).

The proof runs as follows. Consider the following experiment: 
a single spin-$\frac{1}{2}$ particle is prepared 
at time $t_1<t$ in the \mbox{state $\left| A \right\rangle $} (for instance, 
in the eigenstate of the spin 
component in the {\em z} direction with eigenvalue +1), and at time $t_2>t$, 
a measurement is 
performed and the system is found in \mbox{a different state $\left| a \right\rangle $}. 
At \mbox{time {\em t}} we have 
a quantum system both preselected in \mbox{the state $\left| A \right\rangle $} 
and postselected in 
\mbox{the state $\left| a \right\rangle $} \cite{ABL64}. 
\mbox {Whatever $\left| A \right\rangle $ and 
$\left| a \right\rangle $}, there exists a trivial NCHV description 
(compatible with QM) for this 
individual preselected and postselected system \cite{Redhead87p124}. 
Now consider a second spin-$\frac{1}{2}$ 
particle independently prepared at time $t_1<t$ in the state $\left| B \right\rangle $;
at time 
$t_2>t$, a measurement confirms that the second particle is still in the state 
$\left| B \right\rangle $ (for simplicity's sake we suppose the free 
Hamiltonian in {\em t} to be zero). 
Now let us see the quantum description of the composite system. We shall 
use greek letters for the 
states of the composite system and latin letters for the states of each particle. 
At time $t_1<t$, 
the system is prepared in the uncorrelated quantum state
\begin{equation}
\left| {\psi _1} \right\rangle =\left| A \right\rangle \otimes \left| B \right\rangle,
\end{equation}
and at time $t_2>t$, a measurement is performed and the system is found 
in the uncorrelated  state
\begin{equation}
\left| {\psi _2} \right\rangle =\left| a \right\rangle \otimes \left| B \right\rangle.
\end{equation}
Therefore, at time {\em t} we have an individual system both preselected 
in the \mbox{state
$\left| {\psi _1} \right\rangle $} and postselected in the \mbox{state 
$\left| {\psi _2} \right\rangle $}. $\left| A \right\rangle $ and 
$\left| a \right\rangle $
are two different spin states for the first particle and $\left| B \right\rangle $ 
is the same state for 
the second particle in both preselection and postselection. In particular, 
for our argument we suppose
\begin{equation}
\left| a \right\rangle =
{1 \over 3}\left( {\left| A \right\rangle -\sqrt 8\left| {A^\bot } \right\rangle } \right),
\label{af(A)}
\end{equation}
\begin{equation}
\left| {a^\bot } \right\rangle =
{1 \over 3}\left( {\sqrt 8\left| A \right\rangle +\left| {A^\bot } \right\rangle } \right),
\label{af(A)2}
\end{equation}
where $\left\{ {\left| A \right\rangle ,\ \left| {A^\bot } \right\rangle } \right\}$ and 
$\left\{ {\left| a \right\rangle ,\ \left| {a^\bot } \right\rangle } \right\}$ 
are two orthonormal 
bases for the states of the first particle. With election (\ref{af(A)}), the probability of 
postselecting $\left| {\psi _2} \right\rangle $ when 
preselecting $\left| {\psi _1} \right\rangle $ 
is
\begin{equation}
\left| {\left\langle {{\psi _2}} \mathrel{\left | {\vphantom {{\psi _2} {\psi _1}}} \right. 
\kern-\nulldelimiterspace} {{\psi _1}} \right\rangle } \right|^{2}={1 \over 9}.
\label{probmax}
\end{equation}

Consider the three physical quantities represented by the following projection operators
$P_\alpha =\left| \alpha  \right\rangle \left\langle \alpha  \right|$,
$P_{\beta +}=\left| {\beta _+} \right\rangle \left\langle {\beta _+} \right|$,
$P_{\beta -}=\left| {\beta _-} \right\rangle \left\langle {\beta _-} \right|$,
where
\begin{equation}
\left| \alpha  \right\rangle =
\left| {A^\bot } \right\rangle \otimes \left| {B^\bot } \right\rangle,
\end{equation}
\begin{equation}
\left| {\beta _\pm } \right\rangle =
{1 \over 2}\left( {\left| A \right\rangle \otimes \left| {B^\bot } 
\right\rangle \pm \sqrt 3\left| {A^\bot } 
\right\rangle \otimes \left| B \right\rangle } \right).
\end{equation}
The state $\left| {\psi _1} \right\rangle $ is an eigenstate of $P_\alpha $, $P_{\beta +}$
and $P_{\beta -}$ with eigenvalue zero; therefore, a measurement of any of these 
projectors will give, 
with certainty, the value zero. Since we can predict with certainty the result of measuring 
$P_\alpha $, $P_{\beta +}$ and $P_{\beta -}$ at \mbox{time {\em t}}, then, 
following \cite{Redhead87p72}, 
in a NCHV theory at \mbox{time {\em t}} there exist 
three \mbox{{\em elements of reality} \cite{EPR35}} corresponding to the 
three physical quantities $P_\alpha $, $P_{\beta +}$ and $P_{\beta -}$ 
and having a value equal to 
the predicted measurement result, zero in all three cases. 
We will designate these elements of 
reality as
\begin{equation}
v[P_\alpha (t)]=v[P_{\beta +}(t)]=v[P_{\beta -}(t)]=0.
\end{equation}

\sloppy{Consider now the physical quantities represented by the projectors 
\mbox{$P_{\gamma +}=\left| {\gamma _+} \right\rangle \left\langle {\gamma _+} \right|$},
\mbox{$P_{\gamma -}=\left| {\gamma _-} \right\rangle \left\langle {\gamma _-} \right|$},
where}
\begin{equation}
\left| {\gamma _\pm } \right\rangle =
{1 \over {2\sqrt 3}}\left( {\sqrt 8\left| A \right\rangle \otimes 
\left| B \right\rangle +
\left| {A^\bot } \right\rangle \otimes \left| B \right\rangle \mp \sqrt 3
\left| A \right\rangle 
\otimes \left| {B^\bot } \right\rangle } \right),
\end{equation}
or, in the basis (\ref{af(A)}) and (\ref{af(A)2})
\begin{equation}
\left| {\gamma _\pm } \right\rangle =
{1 \over 6}\left[ {3\sqrt 3\left| {a^\bot } \right\rangle \otimes 
\left| B \right\rangle \mp 
\left( {\left| a \right\rangle \otimes \left| {B^\bot } \right\rangle +
\sqrt 8\left| {a^\bot } 
\right\rangle \otimes \left| {B^\bot } \right\rangle } \right)} \right].
\end{equation}
Since $\left| {\psi _2} \right\rangle $ is an eigenstate of $P_{\gamma +}$ and 
$P_{\gamma -}$ with 
zero eigenvalues, then we can infer (retrodict \cite{Vaidman93}), 
with certainty, the result of 
measuring $P_{\gamma +}$ and $P_{\gamma -}$ at \mbox{time {\em t}}; therefore, 
following an extended 
definition for elements of reality proposed by Vaidman \cite{Vaidman93} 
(consisting of the change of 
``predict'' to ``infer'' in Redhead's sufficient condition for elements of reality 
\cite{Redhead87p72}), 
at the \mbox{time {\em t}}, there  
exist two more elements of reality corresponding to these physical 
quantities and having a  value 
equal to the inferred measurement result; that is,
\begin{equation}
v[P_{\gamma +}(t)]=v[P_{\gamma -}(t)]=0.
\end{equation}
Finally, consider the physical quantities  
$P_{\delta +}=\left| {\delta _+} \right\rangle \left\langle {\delta _+} \right|$,
$P_{\delta -}=\left| {\delta _- } \right\rangle \left\langle {\delta _- } \right|$, 
where
\begin{equation}
\left| {\delta _\pm } \right\rangle =
{1 \over {2\sqrt 3}}\left[ {\sqrt 6\left| A \right\rangle \otimes \left| {B^\bot } 
\right\rangle \pm \left( {2\left| A \right\rangle \otimes \left| B \right\rangle -
\sqrt 2\left| {A^\bot } \right\rangle \otimes \left| B \right\rangle } \right)} \right].
\end{equation}
The propositions $P_\alpha $, $P_{\beta +}$, $P_{\gamma +}$, $P_{\delta +}$
form a set of compatible observables for the composite system and, therefore, 
on any  individual 
quantum system, we can measure them jointly without mutual disturbance. In addition, 
they  are 
mutually orthogonal projectors and provide a resolution of the identity 
\begin{equation}
P_\alpha +P_{\beta +}+P_{\gamma +}+P_{\delta +}=I.
\end{equation}
Therefore, in any joint measurement of $P_\alpha $, $P_{\beta +}$, $P_{\gamma +}$,  
$P_{\delta +}$ 
{\em in any state}, the results must be one 1 and three zeros. 
This allows us to use a particular case of  
the {\em sum rule} \cite{Redhead87}: at the \mbox{time {\em t}} the values in a 
NCHV theory must satisfy
\begin{equation}
v[P_\alpha (t)]+v[P_{\beta +}(t)]+v[P_{\gamma +}(t)]+v[P_{\delta +}(t)]=1.
\end{equation}
Since in our preselected and postselected 
individual system $v[P_\alpha (t)]=v[P_{\beta +}(t)]=v[P_{\gamma +}(t)]=0$, 
we are forced to conclude that $v[P_{\delta +}(t)]=1$. Similarly, since $P_\alpha $, 
$P_{\beta -}$, $P_{\gamma -}$, 
$P_{\delta -}$ form another set of compatible observables and a resolution 
of the identity, a 
completely analogous reasoning leads us to conclude that  
$v[P_{\delta -}(t)]=1$. But $P_{\delta +}$ and $P_{\delta -}$ are commutative and 
orthogonal projections representing 
compatible and mutually exclusive physical propositions, so the results of 
any joint measurement 
of $P_{\delta +}$ and $P_{\delta -}$ can never {\em both} be 1. 
So we have reached a contradiction between QM and NCHV for a system preselected and 
postselected in uncorrelated states.

The reason why NCHV models compatible with QM are impossible for 
this composite system (although 
they exist for each particle) is because the dimension of 
the whole quantum system is 4 and, therefore, the 
Gleason-Bell-Kochen-Specker theorem 
applies. In particular, in our argument, a NCHV theory must assign definite 
values to some propositions 
that cannot be measured by {\em local} measurements on each 
particle but only by non-local 
measurements on both particles (in our example, these propositions 
are $P_{\beta +}$, 
$P_{\gamma +}$, $P_{\delta +}$, $P_{\beta -}$, $P_{\gamma -}$, $P_{\delta -}$).
The particular 
election of propositions involved in the argument has been made in order to achieve 
the maximum probability (\ref{probmax}) for the preselection and 
postselection process, preserving the 
relations of orthogonality among states and projectors necessary for the proof.

The same structure of orthogonality relations is behind Hardy's 
proof of Bell's theorem
\cite{Hardy92}
and also appears in some recent proofs of the 
Gleason-Bell-Kochen-Specker
theorem \cite{CG,CEG}. In Hardy's example an individual system 
is preselected in an 
entangled state $\left| {\eta _1} \right\rangle $, which is orthogonal to 
three unentangled states
\begin{equation}
\left| {\hat{\alpha} } \right\rangle =
\left| {A } \right\rangle \otimes \left| {B } \right\rangle,
\end{equation}
\begin{equation}
\left| {\hat{\beta} _+ } \right\rangle=
\left| {a } \right\rangle \otimes \left| {B^\bot } \right\rangle,
\end{equation}
\begin{equation}
\left| {\hat{\beta} _- } \right\rangle=
\left| {A^\bot } \right\rangle \otimes \left| {b } \right\rangle,
\end{equation}
where 
$\left\{ {\left| A \right\rangle ,\ \left| {A^\bot } \right\rangle } \right\}$ and 
$\left\{ {\left| a \right\rangle ,\ \left| {a^\bot } \right\rangle } \right\}$ 
are two orthonormal 
bases for the states of the first particle and
$\left\{ {\left| B \right\rangle ,\ \left| {B^\bot } \right\rangle } \right\}$, and 
$\left\{ {\left| b \right\rangle ,\ \left| {b^\bot } \right\rangle } \right\}$ 
are two orthonormal 
bases for the states of the second particle.
The system is also postselected in the unentangled state
\begin{equation}
\left| {\eta _2} \right\rangle =\left| a \right\rangle \otimes 
\left| b \right\rangle,
\end{equation}
which is orthogonal to
\begin{equation}
\left| {\hat{\gamma} _+ } \right\rangle=
\left| {a^\bot } \right\rangle \otimes \left| {B^\bot } \right\rangle,
\end{equation}
\begin{equation}
\left| {\hat{\gamma} _- } \right\rangle=
\left| {A^\bot } \right\rangle \otimes \left| {b^\bot } \right\rangle.
\end{equation}
Considering also the states
\begin{equation}
\left| {\hat{\delta} _+ } \right\rangle=
\left| {A^\bot } \right\rangle \otimes \left| {B } \right\rangle,
\end{equation}
\begin{equation}
\left| {\hat{\delta} _- } \right\rangle=
\left| {A } \right\rangle \otimes \left| {B^\bot } \right\rangle,
\end{equation}
we have two orthogonal 
resolutions of the identity: 
$\left\{P_{\hat{\alpha}}, P_{\hat{\beta} +}, P_{\hat{\gamma} +}, 
P_{\hat{\delta} +}\right\}$ and
$\left\{P_{\hat{\alpha}}, P_{\hat{\beta} -}, P_{\hat{\gamma} -}, 
P_{\hat{\delta} -}\right\}$. 
Therefore
we have the same relations of orthogonality as in the previous example. 
The connection between these and 
Hardy's proof is explained in \cite{CEG}. For Hardy's example 
the maximum probability for 
the preselection and 
postselection process is \cite{Hardy92} 
\begin{equation}
\left| {\left\langle {{\eta _2}} 
\mathrel{\left | {\vphantom {{\eta _2} {\eta _1}}} \right. 
\kern-\nulldelimiterspace} {{\eta _1}} \right\rangle } \right|^{2}=
{\left( {{{\sqrt 5-1} \over 2}} \right)^5},
\end{equation}
which is smaller than Eq. (\ref{probmax}).  
On the other hand, in Hardy's example all states (except the preselected) 
are unentangled, so it also works as a no-local-hidden-variables proof.

All along in this paper it has been assumed that every 
projector on a Hilbert space 
represents a physical proposition; i.\ e., that there exists an experimental 
setup for measuring it. 
Several results suggest that there is no problem in 
designing such setups, since any discrete 
unitary operator admits an experimental realization in 
terms of optical devices \cite{RZBB94} 
or generalized Stern-Gerlach experiments \cite{SwiftWright80}. 
Therefore, each of the quantum 
inferences used in the argument (predictions, retrodictions, 
and the sum rule for an orthogonal 
resolution of the identity) can be experimentally tested 
(although not all of them on the same 
individual system). 

In summary, a contradiction between QM and NCHV 
models can be found, even for 
a system composed of two 
uncorrelated parts, each of them described by 
two-dimensional Hilbert spaces.\\

I would like to acknowledge Guillermo Garc\'{\i}a Alcaine and
Asher Peres for their many helpful comments.
\pagebreak

\end{document}